\documentclass[12pt,draft]{article}
 
\newcommand{\be}{\begin{equation}}
\newcommand{\ee}{\end{equation}}
\newcommand{\bea}{\begin{eqnarray}}
\newcommand{\eea}{\end{eqnarray}}
\newcommand{\beaa}{\begin{eqnarray*}}
\newcommand{\eeaa}{\end{eqnarray*}}

\newcommand{\del}{\partial}
\newcommand{\g}{{\bf g}}
\newcommand{\h}{{\bf h}}
\newcommand{\p}{{\bf p}}

\newcommand{\J}{{\cal J}}
\newcommand{\K}{{\cal K}}
\newcommand{\La}{{\cal L}}
\newcommand{\Ha}{{\cal H}}

\newcommand{\BB}{{{\rm I} \kern -2pt \rlap {\rm B} \kern +8pt}}

\newcommand{\hl}{\\ \hline \\}

\advance\textheight by \topskip
\makeatletter

\@addtoreset{equation}{section}

\def\section{\@startsection {section}{1}{\z@}{-3.5ex plus -1ex minus
 -.2ex}{2.3ex plus .2ex}{\large\bf\centering}}
\def\subsection{\@startsection{subsection}{2}{\z@}{-3.25ex plus%
 -1ex minus -.2ex}{1.5ex plus .2ex}{\bf}}
\def\subsubsection{\@startsection{subsubsection}{3}{\z@}{-3.25ex plus%
 -1ex minus -.2ex}{1.5ex plus .2ex}{\sl}}
\makeatother

\begin{document}

\baselineskip 18pt
\parindent 12pt
\parskip 10pt

\begin{titlepage}
\begin{flushright}
PUPT-1955\\
DAMTP-2000-124\\
hep-th/0101231\\
December 2000\\[3mm]
\end{flushright}
\vspace{.4cm}
\begin{center}
{\Large {\bf
Integrable Sigma-models and Drinfeld-Sokolov Hierarchies}}\\
\vspace{1cm}
{\large Jonathan M. Evans${}^{a,b}$\footnote{e-mail: 
J.M.Evans@damtp.cam.ac.uk},} 
\\
\vspace{3mm}
{\em ${}^a$ Joseph Henry Laboratories, 
Princeton University, Princeton NJ 08544, U.S.A.}\\
{\em ${}^b$ DAMTP, University of Cambridge, Silver Street, Cambridge
CB3 9EW, U.K.}\\
\end{center}

\vspace{1cm}
\begin{abstract}
\noindent
Local commuting charges in sigma-models with classical Lie groups as 
target manifolds are shown to be related to the conserved quantities 
appearing in the Drinfeld-Sokolov (generalized mKdV) hierarchies. 
Conversely, the Drinfeld-Sokolov construction can be used to deduce
the existence of commuting charges in these and in wider classes of 
sigma-models, including those whose target manifolds are exceptional 
groups or symmetric spaces. This establishes a direct link between 
commuting quantities in integrable sigma-models and in affine Toda 
field theories.

\end{abstract}

\end{titlepage}

\section{Introduction}

Some recent work \cite{EHMM1} established the existence of
infinite families of local, conserved, commuting charges 
in each two-dimensional principal chiral model (PCM) with target 
space a compact, classical Lie group $G$.
The currents underlying these charges are defined using totally symmetric 
$G$-invariant tensors $k^{(m)}_{a_1 a_2 \ldots a_m}$, 
where indices $a$ refer to a basis for the Lie algebra $\g$
corresponding to $G$. The $k$-tensors and their currents are given by 
the formulas
\begin{eqnarray}
\K_m & = & k^{(m)}_{a_1 a_2 \ldots a_m} \, j^{a_1} j^{a_2} \ldots j^{a_m}
\label{keqn} \\
& = & \det ( 1 - \mu j^a t_a )^{s/h} \, \big |_{\mu^{s+1}} \; , \qquad
s = m-1 \, , \ 
\label{deteqn} 
\end{eqnarray}
where $j^a t_a$ is a Noether current 
(arising from a global $G$ symmetry of the model) which takes values
in the defining representation of the classical algebra $\g$ 
with generators $t_a$ (other conventions will be given below).
The tensors $k^{(m)}$ defined by (\ref{deteqn}) 
are non-vanishing precisely when the spin of the 
corresponding charge, $s = m -1$, is equal to an exponent of $G$
modulo its Coxeter number, $h$.

One motivation for the work just mentioned was the 
appearance of certain common features \cite{chari95} 
of exact S-matrices for PCMs \cite{ORW} and affine Toda field 
theories (ATFTs) (see \cite{corri94} and references therein).
The implications of commuting charges with 
spins as given above have been thoroughly studied in the Toda 
case, as reviewed in \cite{corri94}, and the appearance of 
analogous charges in PCMs indeed offers a natural explanation for 
the otherwise mysterious similarities displayed by their $S$-matrices
\cite{EHMM1}.
Subsequently, it was shown that similar families of commuting charges 
can still be constructed if a Wess-Zumino (WZ) term is added to 
the PCM \cite{EHMM2},
or if the target space is some compact symmetric space rather than a 
Lie group \cite{EM}. Analogous results also hold
for supersymmetric models \cite{EHMM2}.
A number of important questions remain unanswered,
however. 

The most obvious problem is to extend the results of 
\cite{EHMM1,EHMM2,EM} to {\it all\/} Lie groups, by including the
exceptional cases along with the classical families.
It is natural to expect that this should be possible,
and yet the formula (\ref{deteqn}) utilizes the defining representation
for each classical algebra,
and so it has no unambiguous interpretation for the exceptional cases.
One can also ask, at a more general level: what is the deeper
mathematical significance of the $k$-tensors, and what is their 
relationship---if any---to more familiar mathematical structures? 
It is reasonable to suppose that an answer to this would help
in passing from classical to exceptional groups. 
Finally, while it was successfully shown in \cite{EHMM1} 
that there are classical commuting charges with identical patterns of spins 
in both PCMs and ATFTs, one may ask whether it is possible to go
further and explicitly relate the charges in the two
models. 

In this paper we shall give answers to each of the three questions 
posed above. 
The next section consists mostly of introductory material on 
sigma-models and their currents, but it finishes with a useful
lemma relevant to the existence of commuting charges.
We then examine, in section 3, whether the formula 
(\ref{deteqn}) gives satisfactory results when applied to 
exceptional groups: it proves entirely adequate for G${}_2$ but 
not for the more complicated cases.
In the remainder of the paper we develop a more systematic 
approach which involves a direct relationship between sigma-model 
and Toda charges; this turns out to be the key to understanding all 
three of the problems mentioned above.

Section 4 summarizes some facts about Toda theories and their
associated Drinfeld Sokolov/modified KdV (DS/mKdV) hierarchies 
of commuting charges \cite{DS1,DS2}. The connection with sigma-models
is particularly natural from the physical perspective
of gauged WZW models \cite{Dublin}, although 
our interest here is to relate PCMs and {\it affine\/} Toda theories,
rather than WZW models and {\it conformal\/} Toda theories.
The links between these various points of view 
will be mentioned below. 

In section 5 we describe an explicit correspondence between 
commuting charges in PCMs on the one hand, 
and DS hierarchies or commuting charges in ATFTs on the other.
This shows that the $k$-tensors of (\ref{deteqn}) are indeed closely-related 
to well-known mathematical structures, and the 
correspondence allows us to deduce the existence of $k$-tensors 
and commuting charges in PCMs based on all Lie groups, including 
the exceptional cases. 
In section 6 we show that similar results can be derived for 
sigma-models based on compact symmetric spaces too.

Appendix A collects together useful data concerning Lie groups 
and symmetric spaces, while appendix B contains some arguments
which supplement remarks made in section 3.

\section{Sigma-models revisited}

We first recall some well-known facts about the classical dynamics and 
canonical structure of a PCM, with or without a WZ term.
It is convenient to use a light-cone canonical 
formalism, where the (real) light-cone coordinates 
$x = x^0 + x^1$ and $\bar x = x^0 - x^1$ play the role of `space' 
and `time' respectively; the corresponding derivatives will be written
$\del =\del_x$ and $\bar \del = \del_{\bar x}$.
A sigma-model with target manifold $G$ can be described in terms of 
a current with components $(j^a , \bar \jmath^a )$ taking values in the Lie
algebra $\g$ and satisfying equations of motion
\be\label{EOMs}
\bar \del j^a = - \del \bar \jmath^a = \kappa f_{bc}{}^a j^b \bar
\jmath^c
\ee 
for some constant $\kappa$.
The structure constants are those appearing in the 
commutation relations $[ t_a , t_b ] = f_{ab}{}^c t_c$
for the generators of $\g$, and all Lie algebra indices 
are raised and lowered using the invariant inner-product $\eta_{ab}$.
(If $G$ is non-compact or if $G$ is compact but our 
basis is not orthonormal then 
the positions of the Lie algebra indices are important.)
 
The equal `time' Poisson brackets can be written 
\be\label{PBs}
\{ j^a (x) , j^b (y) \} = f^{ab}{}_c \, \tilde \jmath^c (x) \, \delta (x{-}y)
\, + \, \eta^{ab}\,  \delta'(x{-}y)
\ee
where $\tilde \jmath^a$ is a certain linear combination of 
$j^a$ and $\bar \jmath^a$. If the coefficient of the WZ term is 
assigned the critical value necessary to define a WZW model then 
$\tilde \jmath^a$ is proportional to $j^a$ alone  
and (\ref{PBs}) becomes a classical Kac-Moody algebra. 
More details concerning the general case can be found in 
{\it e.g.\/} \cite{EHMM2}\footnote{
In \cite{EHMM2} we considered explicitly only compact groups,
and used a canonical formalism with $x^0$ as time. These differences 
entail only very minor modifications of the discussion, however. }

It is a simple consequence of the equations of motion (\ref{EOMs})
that any current defined by (\ref{keqn}) is conserved provided that
$k^{(m)}_{a_1 \ldots a_m}$ is an invariant tensor:
\be\label{kinv}
k^{(m)}_{(a_1 \ldots a_{m-1}}{}^c f^{\vphantom{(a)}}_{b) d c} = 0 
\qquad \Rightarrow 
\qquad  
\bar \del \K_m = 0 \ .
\ee
Now consider the Poisson bracket of two charges
constructed from such currents 
\bea
&&\left \{ \, \int d x \, \K_{m} (x) 
\, , \,
\int d y \, \K_{n} (y) 
\, \right \}
\nonumber \\
&& \quad = \quad 
\int d x 
\int d y \, \, 
k^{(m)}_{a_1 \ldots a_{m}} \, k^{(n)}_{b_1 \ldots b_{n}} \,
\left \{ \, 
j^{a_1} (x) \ldots j^{a_{m}} (x)
\, , \,
j^{b_1} (y) \ldots j^{b_{n}} (y)
\, \right \} \ . \nonumber 
\eea 
In calculating this using (\ref{PBs}),
all contributions involving the term $\delta( x{-}y)$ and the structure 
constants ultimately vanish, by invariance of each $k$-tensor. 
The $\delta'(x{-}y)$ terms contribute a non-trivial integrand, however, and
it is easy to check that this becomes a total derivative, implying
that the charges commute, if and only if
\be\label{kident}
k^{(m)}_{(a_1 \ldots a_{m-1}}{}^c
k^{(n)}_{b_1 \ldots b_{n-2} ) b_{n-1}}{}_c
=
k^{(m)}_{(a_1 \ldots a_{m-1}}{}^c
k^{(n)}_{b_1 \ldots b_{n-2} b_{n-1})}{}_c \ .
\ee
The principal result established in \cite{EHMM1} is that this condition 
holds when the tensors $k^{(m)}$ are defined by the formula 
(\ref{deteqn}) for each compact classical group.\footnote{
Actually, more general families are allowed for the groups 
B${}_n$ and C${}_n$, in which $h$ is replaced by an arbitrary real 
parameter \cite{EHMM1}. 
This possibility will not be of much concern to us here.}
We emphasize that both the conservation of the currents and the commutation 
of their charges are completely independent of the presence or 
absence of a WZ term in the model.

From this point on we will take $G$ to be compact unless we 
explicitly state otherwise (we will discuss briefly in the next
section some aspects of hamiltonian reduction, for which $G$ is required
to be maximally non-compact). Our aim in the remainder of this section
is to show that the validity, or otherwise, of the condition (\ref{kident}),
and hence the question of commuting charges, can be settled 
essentially by restricting attention to a Cartan subalgebra.
Although the {\em restriction lemma\/} which we shall establish is 
quite simple, it will prove very useful throughout the remainder 
of the paper.
To explain the arguments properly, we first need to recall 
some standard results \cite{Helg,Adams} 
and introduce some notation.

Let $\g$ be a compact Lie algebra and $\g_0$ a Cartan subalgebra
(CSA). We assume, with no loss of generality, that our basis 
$\{ t_a \}$ for $\g$ can be partitioned into bases $\{ t_i \}$ for $\g_0$
and $\{ t_\alpha \}$ for its orthogonal complement $\g_0^\perp$.
Allowing complex linear combinations of generators,
the latter may be chosen to consist of the usual Cartan-Weyl 
step operators corresponding to the non-zero roots of $\g$.
Recall that any $X = X^at_a \in \g$ is conjugate to some member 
of our chosen CSA, {\it i.e.\/} there exists $g \in G$ such that
$gXg^{-1} \in \g_0$. The remnant of $G$ 
which fixes the Cartan subalgebra (under the adjoint action) 
is the Weyl group, ${\bf W}(G)$. 

For a tensor $d^{(m)}$ of degree $m$ on $\g$, and any 
$U, V, \ldots , Z \in \g$ we write
$$
d^{(m)} (U, V,  \ldots , Z) = d^{(m)}_{a_1 a_2 \ldots a_m}
U^{a_1} V^{a_2} \! \ldots \, Z^{a_m}
$$
so that the components of the tensor can be expressed 
\be\label{kcomp} 
d^{(m)}_{a_1 a_2 \ldots a_m} = d^{(m)} ( t_{a_1} , t_{a_2} , \ldots ,
t_{a_m} )
\ee
(we shall not always indicate the degree of the tensor explicitly).
Such a tensor is $G$-invariant if 
\be\label{infinv}
d ( \, [T, U ] \, , \, V \, , \ldots , \, Z \, ) \, + \, 
d ( \, U \, , \, [T, V ] \, , \ldots , \, Z \, ) \, + 
\ldots
+ \, d ( \, U \, , \, V \, , \ldots , \, [T , Z] \, ) = 0 
\ee
for all $T \in \g$
(which coincides with the condition written 
earlier in (\ref{kinv})); 
or equivalently 
\be\label{fininv}
d ( \, gUg^{-1} , \, gVg^{-1} , \ldots , \, gZg^{-1} ) = 
d (U, V , \ldots , Z) 
\ee
for all $g \in G$.
If $d$ is totally symmetric, then it is completely determined 
by specifying $d (X, \ldots , X)$ 
for all $X \in \g$.
From our earlier remarks, it follows that any symmetric invariant 
tensor is determined by its restriction to the CSA, this 
restricted tensor having components 
\be 
d^{(m)}_{i_1 i_2 \ldots i_m} = d^{(m)} ( t_{i_1} , t_{i_2} , \ldots ,
t_{i_m} ) 
\ee 
The restricted tensor is invariant under the Weyl group ${\bf W}(G)$.

A useful observation is that any symmetric invariant tensor
satisfies 
\be\label{zeroabs} 
d_{\alpha \, i_1 \ldots i_{m-1}} = 0 
\qquad {\rm or} \qquad 
d ( Y, X , \ldots , X ) = 0 \qquad {\rm for} \quad X \in \g_0 \ , \quad
Y \in \g_0^\perp \ .
\ee
This can be understood in a number of ways---
for example, it is a consequence of the 
usual ${\bf Z}$-grading of $\g$ defined by the choice 
of CSA and a set of simple roots: 
a particular component of an invariant tensor must vanish unless the 
total grade associated with all the indices it carries is zero.\footnote{
The grade of any element of $\g$ is its eigenvalue under commutation
with a specific element $M \in \g_0$; 
all members of the CSA therefore have grade zero,
while step operators for the positive/negative simple roots are
assigned grades $\pm 1$, by construction. 
The result follows from (\ref{infinv}) with $T=M$}.
Alternatively, consider some specific $Y = t_\alpha \in \g_0^\perp$ 
corresponding to a non-zero root of $\g$. 
For each such element, we can choose $T = t_i \in \g_0$ (so    
$[T, X] = 0$) such that $[T, Y] = \lambda Y$ with $\lambda \neq 0$. 
The invariance condition 
(\ref{infinv}) with $U=Y$ and $V = \ldots = Z = X$ 
implies that (\ref{zeroabs}) holds for this
particular $Y$, and the result follows for general $Y \in \g_0^\perp$ 
by linearity. 

Following these remarks on general symmetric invariant tensors,
we now focus specifically on the property necessary 
to construct commuting charges, and establish the following. 
\hfil \break 
\noindent
{\em Restriction lemma}:
totally symmetric, $G$-invariant tensors 
$k^{(m)}$ and $k^{(n)}$ on $\g$ satisfy the condition 
(\ref{kident}) if and only if their restrictions to the CSA $\g_0$ 
obey the analogous condition:
\be\label{csaident}
k^{(m)}_{(i_1 \ldots i_{m-1}}{}^\ell \,
k^{(n)}_{j_1 \ldots j_{n-2} ) j_{n-1}}{}_\ell
=
k^{(m)}_{(i_1 \ldots i_{m-1}}{}^\ell \,
k^{(n)}_{j_1 \ldots j_{n-1})}{}_\ell \ .
\ee
where $\ell$ is a CSA index. 
\hfil \break
{\em Proof:} (\ref{kident}) holds iff  
\bea 
&&
k^{(m)}_{a_1 \ldots a_{m-1}}{}^c \,
k^{(n)}_{b_1 \ldots b_{n-2}  b_{n-1}}{}_c \, 
X^{a_1} \! \ldots X^{a_{m-1}} 
X^{b_1} \! \ldots X^{b_{n-2}} Y^{b_{n-1}}
\nonumber \\
&= & 
k^{(m)}_{(a_1 \ldots a_{m-1}}{}^c \, 
k^{(n)}_{b_1 \ldots b_{n-2}  b_{n-1})}{}_c \, 
X^{a_1} \! \ldots X^{a_{m-1}} 
X^{b_1} \! \ldots X^{b_{n-2}} Y^{b_{n-1}}
\nonumber
\eea
for any $X, Y \in \g$.
Because the tensors are invariant, this is equivalent to 
\bea 
&& k^{(m)}_{i_1 \ldots i_{m-1}}{}^c \, 
k^{(n)}_{j_1 \ldots j_{n-2}  b_{n-1}}{}_c \,
X^{i_1} \! \ldots X^{i_{m-1}}
X^{j_1} \! \ldots X^{j_{n-2}} Y^{b_{n-1}}
\nonumber \\
& = &
k^{(m)}_{(i_1 \ldots i_{m-1}}{}^c \,
k^{(n)}_{j_1 \ldots j_{n-2}  b_{n-1})}{}_c \, 
X^{i_1} \! \ldots X^{i_{m-1}}
X^{j_1} \! \ldots X^{j_{n-2}} Y^{b_{n-1}}
\nonumber
\eea
with $X \in \g_0$ and $Y \in \g$.
Now (\ref{zeroabs}) 
implies that one or other of the tensor factors on each side of this
equation will vanish unless $Y \in \g_0$ and the summation over
$c$ is also restricted to the CSA. Hence, the original condition 
holds iff 
\bea 
&&
k^{(m)}_{i_1 \ldots i_{m-1}}{}^\ell \, 
k^{(n)}_{j_1 \ldots j_{n-2}  j_{n-1}}{}_\ell \,
X^{i_1} \! \ldots X^{i_{m-1}}
X^{j_1} \! \ldots X^{j_{n-2}} Y^{j_{n-1}}
\nonumber \\
& = &
k^{(m)}_{(i_1 \ldots i_{m-1}}{}^\ell \,  
k^{(n)}_{j_1 \ldots j_{n-2}  j_{n-1})}{}_\ell \, 
X^{i_1} \! \ldots X^{i_{m-1}}
X^{j_1} \! \ldots X^{j_{n-2}} Y^{j_{n-1}}
\nonumber
\eea 
for all $X, Y \in \g_0$. This is equivalent to (\ref{csaident}),
completing the proof.

\section{Exceptional groups: a direct approach}

In this section we investigate whether the formula  
(\ref{deteqn}) might be of use when applied to exceptional groups.
To formulate this question properly, one must first choose a 
representation for the group to which the generators appearing 
in (\ref{deteqn}) will belong. 
To answer it, one must then determine whether the 
resulting currents $\K_{s+1}$ share the properties of
their counterparts for the classical
algebras: 
(i) that they vanish identically unless $s$, the spin of the 
conserved charge, is equal to an exponent of the algebra modulo
its Coxeter number $h$; (ii) that these conserved charges commute.

It is not obvious how to carry out calculations for an 
exceptional group $G$ in the same manner as was done for the 
classical families in \cite{EHMM1}. We can, however, make
use of these earlier results by combining them with the restriction lemma 
proved at the end of the last section. This lemma tells us that
it is sufficient to calculate the quantities $\K_m$ and their Poisson 
brackets when the underlying Lie algebra variables $j^a t_a$  
are restricted to a CSA of $\g$. More conveniently, we 
can choose a classical subgroup of maximal rank $H \subset G$,
with Lie subalgebra $\h \subset \g$,
and carry out all calculations assuming $j^a t_a$ belongs to
$\h$. Because $\h$ contains a CSA of $\g$, the restriction lemma
ensures that these results will reveal all the information we 
seek. We must, of course, take into account that 
our chosen representation of $G$ will in general decompose 
into various irreducible representations of $H$.

Of the five exceptional groups, ${\rm G}_2$ is certainly the simplest 
and also the one which most nearly possesses something like a defining 
representation,\footnote{The
appropriate definition of ${\rm G}_2$ is the automorphism group of the
octonions, with the 7-dimensional space of pure-imaginary octonions 
furnishing the representation \cite{Adams}.}
of dimension 7.
A classical subgroup of maximal rank is ${\rm SU(3)} \subset {\rm G}_2$,
with respect to which this representation decomposes 
${\bf 7} = {\bf 3} \oplus {\bf 3}^* \! \oplus {\bf 1}$.
Following the strategy explained above, we consider a 
current $j^a t_a$ belonging to the ${\rm SU}(3)$ subalgebra.
Let us introduce the notation\footnote{
In the notation of \cite{EHMM1}:
$2a = \J_2 = {\rm Tr} (j^2)$ and $3b = \J_3 = {\rm Tr} (j^3)$
for $j = j^a t_a$ an SU(3) current.}
\bea
A(x, \mu) & = & \det ( 1 - \mu j^a t_a) 
\qquad {\rm for} \ j^at_a \ {\rm in}~{\bf 3}~{\rm of} \ {\rm SU(3)}\, ,
\nonumber \\
& = & 1 - \mu^2 a(x) - \mu^3 b(x) 
\eea
which defines $A(x,\mu)$, $a(x)$ and $b(x)$; consequently 
\bea
A(x, -\mu) & = & \det ( 1 - \mu j^a t_a) 
\qquad {\rm for} \ j^at_a \ {\rm in}~{\bf 3}^*~{\rm of} \ {\rm SU(3)}
\, ,
\nonumber \\
& = & 1 - \mu^2 a(x) + \mu^3 b(x) \ . 
\eea
Now define 
\bea
B(x ,\mu) & = & \det ( 1 - \mu j^a t_a) 
\qquad {\rm for} \ j^at_a \ {\rm in}~{\bf 7}~{\rm of} \ {\rm G}_2 \, , 
\nonumber \\ 
& = & A(x, \mu) A(x, -\mu) 
\nonumber \\
& = & (1 - \mu^2 a )^2 - \mu^6 b^2  
\label{Beqn} \eea
where the second equality follows from the decomposition of 
representations given above.
The Coxeter number for ${\rm G}_2$ is $h=6$, and so the formula 
(\ref{deteqn}) becomes  
\be
\K_{s+1} = B(x, \mu)^{s/6} \, \Big |_{\mu^{s+1}}
\ee
where the expansion is to be taken in ascending powers of $\mu$.
Now let us consider whether these currents have the desired properties.

Since $B(x,\mu)$ is an even polynomial in $\mu$, $\K_{s+1}$ 
is non-zero only if $s$ is an odd integer. 
In order to investigate in more detail which of these expressions are 
non-vanishing, it is convenient to write   
\bea 
B(x, \mu)^{s/6} & = & (1 - \mu^2 a)^{s/3} \left [
1 - \mu^6 b^2 (1 - \mu^2 a)^{-2} \right ]^{s/6}
\nonumber \\
& = & (1 - \mu^2 a)^{s/3} 
\sum_{p\geq 0} \, c_p \, \mu^{6p} \, b^{2p} \, 
(1 - \mu^2 a)^{-2p}
\label{Bexp}
\eea
where $c_p$ are certain binomial coefficients, and the brackets 
appearing in each term of the sum are yet to be expanded as power series in
$\mu$. Now, if $s = 6n+3$ for some integer $n$, then 
\be
B(x, \mu)^{s/6} = \sum_{p\geq 0} \, c_p \, \mu^{6p} \, b^{2p} \, 
(1 - \mu^2 a)^{2n+1-2p} \ , 
\ee
and $\K_{s+1}$ is, by definition, the coefficient of $\mu^{s+1} = \mu^{6n+4}$.
Because of the factor of $\mu^{6p}$ in this sum,
those terms with $p>n$ produce powers of $\mu$ which are never less
than $6n+6$, and so never contribute to $\K_{s+1}$.
On the other hand, for terms with $p \leq n$ the power of 
the bracket $(1-\mu^2 a)$ is a positive integer, and the highest
power of $\mu$ which arises after its expansion is therefore 
$2(2n+1-2p) + 6p = 4n+2p+2 < 6n+4$, so these terms cannot contribute either. 
We conclude that $\K_{s+1}$ vanishes for $s=6n+3$, while in general it
will be non-vanishing for $s=6n+1$ or $s= 6n+5$. This is precisely
what we require, because the exponents of ${\rm G}_2$ are 1 and 5.
The formula (\ref{deteqn}) for the {\bf 7} representation of ${\rm G}_2$ 
has therefore passed the first test.

The second test, that the non-trivial conserved charges commute,
is even more stringent.
We will now calculate the relevant Poisson brackets using 
similar methods to those of \cite{EHMM1,Dublin}. 
It was shown in \cite{EHMM1} (equation (4.14))
that the function $A(x,\mu)$ introduced above has Poisson brackets
\bea 
\{ A(x, \mu) , A(y, \nu) \} \; = \; 
\mu^2 \nu^2  
\left [ \, {\del_\mu{-}\del_\nu \over \mu{-}\nu} +
{\del_\mu \del_\nu \over 3} \, \right ] 
A(x,\mu) \, \del_x ( \, A(x,\nu) \, \delta(x{-}y) \, )
\phantom{XXXX} \nonumber \\
+ \; 
{\mu^2 \nu^2 \over (\mu{-}\nu)^2} 
\left [ \, \del_x A(x,\mu) \, A(x,\nu) - A(x,\mu) \, \del_x A(x,\nu) \,
\right ]
\delta (x{-}y) \, . 
\qquad
\label{APB}
\eea
From this we can calculate 
\bea 
\{ B(x, \mu) , B(y, \nu) \} \; = \; 
{2 \mu^2 \nu^2 \over \mu^2{-}\nu^2} 
\left [ \; \vphantom{{1\over2}} 
( \mu \del_\mu - \nu \del_\nu ) B(x,\mu) \, \del_x (\, B(x,\nu) \, 
\delta(x{-}y) \, )
\right . 
\phantom{XXXXXX}
\nonumber \\
\phantom{XXXXXXXXXX} + \; \left . {\mu^2{+}\nu^2 \over \mu^2{-}\nu^2} \,
( \, \del_x B(x,\mu) \, B(x,\nu) - B(x,\mu) \del_x B(x,\nu) \, ) \, 
\delta (x{-}y)
\; \right ]
\nonumber 
\\[2pt]
+  \; {\mu^2 \nu^2 \over 3} \, C(x,\mu) \, \del_x ( \, C(x,\nu) 
\delta (x{-}y) \, )
\phantom{XXXXXXXXXX} 
\label{BPB}
\eea
where
\bea
C(x, \mu) & = & 
\del_\mu A(x, \mu) \, A(x,-\mu) \, - \, A(x,\mu) \, \del_\mu A(x,-\mu) 
\nonumber \\
& = & - 2 b \mu^2 (3 - a \mu^2) \ .
\label{Ceqn}
\eea
With the exception of the additional $C$-terms, the Poisson bracket
(\ref{BPB}) is again familiar from \cite{EHMM1} 
(equation (4.25) with $\mu$ and $\nu$ replaced 
by $\mu^2$ and $\nu^2$ respectively).

To find the Poisson brackets of two conserved charges, we must 
evaluate
\be
\int \! dx \, \int \! dy \, \, \{ B(x,\mu)^{s/6} , B(y,\nu)^{r/6} \}
\ee 
in sufficient detail to extract the coefficient of $\mu^{s+1} \, \nu^{r+1}$.
Following precisely the same arguments as in section 4.3 of 
\cite{EHMM1}, it can be shown that the $B$-terms 
on the right-hand-side of (\ref{BPB}) will not contribute to this
result. The additional $C$-terms, however, produce an
expression proportional to
\be
\int \! dx \, 
\mu^2 \nu^2 \, 
B(x,\mu)^{s/6-1} C(x,\mu) \, \del_x ( \, B(x,\nu)^{r/6-1} C(x,\nu) ) \
.
\ee
Using (\ref{Ceqn}), and extracting the relevant powers 
of $\mu$ and $\nu$, we find that the 
integrand contains a factor 
\be
3 B(x, \mu)^{(s-6)/6} \Big |_{\mu^{s-3}} - a(x)  
B(x, \mu)^{(s-6)/6} \Big |_{\mu^{s-5}} \ .
\ee
But this can be shown to vanish by 
expanding each term in the form (\ref{Bexp}) (with $s$ replaced 
by $s-6$) and noting that  
\be 
(1 - a \mu^2)^{(2n-1)/3} \Big |_{\mu^{2n+2}}
=
{a \over 3} \, (1 - a \mu^2)^{(2n-1)/3} \Big |_{\mu^{2n}}
\ee
for any integer $n$ (if the general term in each expansion is 
labelled by $p$, we have set $s-5 - 6p= 2n$).
Hence, the charges commute.

We have shown that the formula
(\ref{deteqn}), taking the seven-dimensional representation of ${\rm G}_2$,
defines currents with all the properties we require.
The remaining exceptional groups can be investigated similarly.
For example, there are convenient classical subgroups of maximal rank:
${\rm SO(9)} \subset {\rm F}_4$ and ${\rm SO(16)} \subset {\rm E}_8$.
In these two instances, the representations of smallest dimension 
and their decompositions 
are ${\bf 26} = {\bf 9} \oplus {\bf 16} \oplus {\bf 1}$ and
${\bf 248} = {\bf 120} \oplus {\bf 128}$ respectively.
In both these cases, however, the formula (\ref{deteqn}) fails
the first test, because it is easy to check that 
there are non-trivial currents $\K_{s+1}$ 
for which $s$ is not congruent (mod $h$)
to an exponent. Furthermore, explicit computations 
for ${\rm F}_4$ show that the currents $\K_6$ and $\K_8$ obtained 
from this formula do not yield commuting charges.

In summary, although the formula (\ref{deteqn}) works perfectly for
${\rm G}_2$, the same cannot be said for the other 
exceptional groups. It would be interesting to investigate 
other representations of these groups, or possible 
modifications of the formula, but these are not issues that we 
shall pursue here.
Some insight into the special nature of ${\rm G}_2$ can be gained by
regarding it as the subgroup of SO(8) invariant under outer
automorphisms. By exploiting the fact that these 
automorphisms become inner when SO(8) is embedded in ${\rm F}_4$,
we can even use the ${\rm G}_2$ tensors defined above to
construct commuting quantities in the ${\rm F}_4$ model. 
Since these arguments lie somewhat outside the main development 
of ideas in this paper, we shall present them in appendix B.
It is, in any case, convenient to delay such discussion
until after we have explained the connection between  
sigma-models and Toda theories,
which will lead ultimately to a more uniform understanding of both
the classical and exceptional cases.

\section{Toda theories and DS/mKdV hierarchies}

We shall be concerned with conformal Toda field theories (CTFTs) 
based on finite dimensional Lie algebras $\g$, and affine Toda field 
theories (ATFTs) associated with (untwisted) affine Kac-Moody algebras
$\widehat \g$. In either case, the Toda fields $\phi^i$ take values 
in the CSA $\g_0 \subset \g$. 
They obey classical equations of motion of the form
\be 
\bar \del \del \phi^i = m^2 \sum_{\alpha \in R} 
\alpha^i \exp (\alpha \cdot \phi )
\ee
where $R$ is a certain subset of the roots (a dot denotes 
the inner-product $\eta_{ij}$ on the CSA).
For a CTFT, $R$ is precisely the set of simple roots of $\g$,
while for an ATFT, $R$ contains in addition 
the lowest root of $\g$ (these are the simple 
roots of $\widehat \g$ projected onto the CSA of the `horizontal'
subalgebra $\g$). The conformal symmetry present in the former case 
means that the mass parameter $m$ can be removed by shifting the
fields, but for an ATFT the effect of the additional term involving
the lowest root is to produce a minimum in the potential, 
resulting in a massive theory.
We use the same light-cone canonical formalism as before, 
with $\bar x$ as `time' and $x$ as `space'. The canonical 
Poisson brackets are identical in both CTFT and ATFT:  
introducing the variables $u^i = \del \phi^i$, the brackets
are 
\be\label{todaPBs} 
\{ u^i (x) , u^j (y) \} = \eta^{ij} \delta' (x{-}y) \ .
\ee

Each CTFT contains an infinite number of conserved
currents which are differential polynomials in the 
quantities $u^i$. They take the form 
\be\label{wtoda} 
{\cal W}_{m} = d_{i_1 i_2 \ldots i_m} u^{i_1} u^{i_2} \ldots u^{i_m} +  ({\rm
derivative~terms}) 
\qquad {\rm with} \qquad 
\bar \del {\cal W}_m = 0
\ee
where the `derivative terms' are lower-order in the fields $u$ 
but may also involve $\del u$, $\del^2 u$, and so on.
Under the Poisson bracket, these conformal currents form a classical 
W-algebra. Much progress in understanding the W-algebra structure of CTFTs
has come from regarding them as constrained WZW models \cite{Dublin},
and we now recall in outline how this is done.

The construction starts with a WZW model based on a maximally
non-compact group $G$. Its Lie algebra is the real span of a 
set of Cartan-Weyl generators, and it
can be decomposed according to the associated integer grading: 
$\g = \g_+ \oplus \g_0 \oplus \g_-$, where $\g_\pm$ are 
the nilpotent subalgebras consisting of elements of positive/negative 
grade respectively. 
The corresponding nilpotent subgroups of $G$ will be denoted by $G_\pm$.
The currents $j$ in the WZW model satisfy $\bar \del j = 0$ 
and obey a Kac-Moody algebra. 

The WZW model is then modified by the addition 
of gauge fields so as to ensure invariance 
under an enlargement of the Kac-Moody symmetry,
corresponding to gauging the nilpotent subgroups $G_\pm$.
As in any gauge theory, only quantities invariant under the 
gauge transformations now have intrinsic meaning.   
A particularly important class of such objects are the 
gauge-invariant differential polynomials in the Kac-Moody currents, 
which take the form
\be\label{diffpoly}
d_{a_1 a_2 \ldots a_m} \, j^{a_1} j^{a_2} \! \ldots j^{a_m} 
\, + \, ({\rm derivative~terms})
\ee
It is shown in \cite{Dublin} that the tensor $d_{a_1 a_2 \ldots a_m}$ must
be $G$-invariant (even though we are considering gauge
transformations involving just the nilpotent subgroups $G_\pm$).
The additional derivative terms have a complicated structure which is 
also dictated by gauge invariance.

The Toda description of such a modified WZW model emerges on making a 
particular gauge choice which allows the sigma-model 
target-space to be parameterized by fields $\phi^i$ living in the CSA.
In terms of these Toda fields, the expressions (\ref{diffpoly}) reduce
to the currents (\ref{wtoda}). An important consequence of this 
is that the tensor $d_{i_1 i_2 \ldots i_m}$ is invariant under the Weyl 
group of $G$.

The holomorphic currents (\ref{wtoda}) can also be constructed 
directly in the CTFT.
An elegant approach is to introduce a (pseudo)differential Lax operator 
$\La (u, \del)$, of order $n$ say, 
which can be expanded in descending powers of derivatives:
\be
\La = \sum_{m \geq 0} \, {\cal W}_m \, \del^{n-m}  \qquad ({\cal W}_0
= 1) \ .
\ee
The operator is constructed so as to obey
\be
[ \bar \del , \La ] = 0 \qquad \Rightarrow \qquad
\bar \del {\cal W}_m = 0 
\ee
and hence the coefficients in its expansion yield the desired 
conserved currents.
For CTFTs based on certain algebras 
there is a particularly simple formula for the Lax operator \cite{MS}:
\be\label{conflax}
\La = \prod_{\lambda \in W} (\del + \lambda \cdot u) 
\qquad {\rm for} \quad {\rm A}_r, \ {\rm B}_r, \ {\rm C}_r \quad {\rm
and} \quad {\rm G}_2 \ ,
\ee
where $W$ is the set of weights of the defining
representation, or the seven-dimensional representation for ${\rm G}_2$,
and the product is taken in the order lowest to highest weights 
from left to right. 
For other algebras, such an ordering is ambiguous and this
complication manifests itself through the appearance of inverse powers
of $\del$. Thus, we have \cite{MS}
\be\label{dconflax}
\La = \prod_{\lambda \in W_-} (\del + \lambda \cdot u) \, {1 \over \del} 
\prod_{\lambda \in W_+ }(\del + \lambda \cdot u) \qquad {\rm for} \quad
{\rm D}_r \ ,
\ee
where $W_\pm$ are the strictly positive/negative weights of the fundamental 
representation, and each product is again ordered 
lowest to highest.
Similar Lax operators have apparently not been calculated for the 
remaining exceptional algebras, although there is no 
obstacle to doing so in principle \cite{MS,Dublin}.

We have now outlined two approaches to the construction of the
infinite set of conserved currents in each TCFT.
An important {\it finite\/} subset of these currents consists of the 
conformal primary fields for the W-algebra \cite{MS,Dublin}. These 
currents have spins $s+1$ where $s$ is an exponent 
of the Lie algebra $\g$.
Another important set of currents, less familiar from the point of view 
of conformal field theory perhaps, are those which give rise to 
commuting charges. It is known from the work of Drinfeld
and Sokolov \cite{DS1,DS2} that there are infinitely many such
currents; from our point of view these correspond to very special 
choices of the $d$-tensors in (\ref{wtoda}). Let us denote these currents 
\be\label{mkdvcurr} 
\Ha_m = h_{i_1 i_2 \ldots i_m} u^{i_1} u^{i_2} \ldots u^{i_m}  
\, + \, ({\rm derivative~terms}) \ .
\ee
Drinfeld and Sokolov establish the existence of a maximal set of 
commuting charges 
\be\label{mkdvcharge} 
H_s = \int d x \, \Ha_{s+1}
\ee
where the spins $s$ take values equal to the exponents of 
$\g$ modulo the Coxeter number $h$. 
Note that there is a finite subset of commuting charges 
whose spins are exactly the exponents, but their currents 
do not coincide, in general, with the primary field currents 
of the W-algebra.

Thus far we have discussed conserved quantities in CTFT
corresponding to a finite-dimensional algebra $\g$.
The ATFT based on $\widehat \g$ is obviously closely 
related: it has the same field content, and the classical equations of 
motion differ only by the addition of one term involving the lowest
root. This extra term has the dramatic consequence of making the ATFT 
a massive theory and so the conformal
conservation laws (\ref{wtoda}) of the CTFT will not survive
in general. There are still infinitely many 
conserved quantities in each ATFT, however, and they involve precisely
the quantities $\Ha_m$ introduced in (\ref{mkdvcurr}) above.
In ATFT these satisfy the modified conservation equations
\be\label{atftcons}
\bar \del \, \Ha_m + \del \, \bar \Ha_m = 0
\ee
for certain $\bar \Ha_m$ which will be complicated 
functions of the fields involving, in particular, exponentials of the 
lowest root. Fortunately, the exact expressions need not concern us,
because in the light-cone canonical formalism $\bar \Ha_m$ 
plays the role of the `space' component of the current,
and hence the conserved charge constructed from (\ref{atftcons})
is given by precisely the same formulas   
(\ref{mkdvcharge}) and (\ref{mkdvcurr}) as before.

To summarize: the set of conserved charges in $\widehat \g$ ATFT 
commute with one another, and they can be identified in a direct way
with the maximal set of commuting charges contained within the W-algebra 
of $\g$ CTFT. The charges $H_s$ regarded as functions of 
the fields $u$ via (\ref{mkdvcharge}) and (\ref{mkdvcurr}) 
constitute precisely the Drinfeld-Sokolov generalizations of 
the mKdV hierarchy (mKdV being the simplest example, associated to the
algebra A${}_1$). For summaries of this and much 
related material see {\it e.g.} \cite{FF}.

Formulas for many of the DS currents $\Ha_m$
were derived in the original works \cite{DS1,DS2}
and they can be expressed in terms of the Lax operators 
$\La$ for CTFT given in (\ref{conflax}) and (\ref{dconflax}).
For each ATFT based on a classical group or on ${\rm G}_2$, we 
introduce a related Lax operator 
$\widehat \La (u , \del)$ of order $h$ (the Coxeter number of 
the algebra) in terms of which 
\be\label{laxdef}
\Ha_{s+1} = {\rm Res} ( \widehat \La^{s/h} ) \ .
\ee
The fractional power must be defined by means of an expansion in 
{\em descending\/} powers of the operator $\del$, and Res is 
an instruction to extract the {\em residue\/}, meaning the 
coefficient of $\del^{-1}$. (For background on such 
techniques for general pseudo-differential operators,
see {\it e.g.} \cite{Dickey}). 
Specifically, we have for each algebra 
\be\label{afflax} 
\widehat \La = \La \, \quad 
{\rm for} \ \ {\rm A}_r \, , \ {\rm C}_r \, ; \qquad \qquad 
\widehat \La = \La \, \del^{-1} \, \quad
{\rm for} \ \ {\rm B}_r \, , \ 
{\rm D}_r \, , \ {\rm G}_2 \, .
\ee
These definitions, in conjunction with (\ref{laxdef}), provide 
concrete expressions for all the commuting charges which arise
in these models, with the exception of those associated with 
the Pfaffian-type invariants of ${\rm D}_r$. (Analogous expressions for 
these currents are apparently not known---see \cite{DS1,DS2}.) 

We conclude this section with one simple deduction from 
the beautiful results of Drinfeld and Sokolov.
With the currents written in the form (\ref{mkdvcurr}),
let us consider the leading term, involving the largest number of 
fields $u$ and the lowest number of derivatives, 
which arises when we calculate directly the Poisson bracket
$\{ H_{m-1} , H_{n-1} \}$ using (\ref{todaPBs}). Since we know this 
bracket vanishes, the entire integrand must be a total derivative, and
its leading term must be a total derivative by itself (it is the
unique term containing the largest possible number of fields $u$).
It is straightforward to see that this requires 
\be\label{hident}
h^{(m)}_{(i_1 \ldots i_{m-1}}{}^\ell \,
h^{(n)}_{j_1 \ldots j_{n-2} ) j_{n-1}}{}_\ell
=
h^{(m)}_{(i_1 \ldots i_{m-1}}{}^\ell \,
h^{(n)}_{j_1 \ldots j_{n-2} j_{n-1})}{}_\ell \ .
\ee 
This is just the condition (\ref{csaident}) encountered earlier.
We shall make the connection explicit in the next section.

\section{DS hierarchies and PCM/WZW models} 

In each PCM or WZW model, we have conserved commuting charges 
based on $G$-invariant $k$-tensors satisfying (\ref{kident}). 
We showed that this condition is valid iff it holds when the tensors
are restricted to the Cartan subalgebra. From direct calculation,
the condition is known to be satisfied when the $k$-tensors are defined by 
(\ref{deteqn}) for each of the classical algebras (in their defining
representations) and for ${\rm G}_2$ 
(in its seven-dimensional representation).
In conformal or affine Toda theory, on the other hand,
we have a set of mutually commuting charges given by the Drinfeld-Sokolov 
construction for any Lie algebra $\g$. 
The $h$-tensors on the CSA which appear in (\ref{mkdvcurr})
are Weyl-invariant, and must satisfy (\ref{hident}) in order that 
the corresponding charges commute. 
 
There is an obvious way to relate the two pictures: extend 
the $h$-tensors of the DS hierarchies to tensors 
$h_{a_1 a_2 \ldots a_m}$ on each compact Lie algebra $\g$. 
Specifically, for any $X \in \g$, we choose $g \in G$ such that 
$gXg^{-1} \in \g_0$ and set 
\be 
h( X , \ldots , X) = h( gXg^{-1} , \ldots , gXg^{-1} )  
\ee
to define a totally symmetric extension of the tensor from $\g_0$ to $\g$.
This extension is unambiguous because the choice of $g$ is unique up to
elements which fix the CSA, but the (adjoint) actions of such
elements on $\g_0$ constitute the Weyl group, under which 
each $h$-tensor is invariant. 
The extended $h$-tensor is $G$-invariant on $\g$, by construction.
By the restriction lemma, (\ref{hident}) is sufficient
to ensure that our extended tensors can be used to define 
commuting charges in the PCM based on $G$, with currents
$h_{a_1 a_2 \ldots a_m} j^{a_1} j^{a_2} \ldots j^{a_m}$.

In this manner, the DS/mKdV hierarchies  
ensure the existence of commuting sets of charges 
in any sigma-model with target space a compact Lie group $G$,
whether classical or exceptional.
It remains to reconcile this new definition 
with our old definition, in terms of 
$k$-tensors given by (\ref{deteqn}), however.
To achieve this we must show that any concrete expressions 
which are available for both the $k$-tensors and $h$-tensors agree
up to irrelevant overall constants.

In the formulas (\ref{laxdef}) 
the differential operators
$\del$ in every factor of the Lax operator 
generate a large number of terms.
But to extract the $h$-tensor for each current we are concerned only 
with the leading, non-derivative terms, as written in
(\ref{mkdvcurr}). Discarding the derivative terms is equivalent 
to neglecting the action of each operator $\del$ on all fields $u$
standing to its right. This can be achieved simply by replacing 
$\del$ in the definition by a parameter $1/\mu$, say, so that 
\be
h_{i_1 i_2 \ldots i_n} u^{i_1} u^{i_2} \ldots u^{i_n} =
\widehat \La (u, 1/\mu)^{s/h} \, \Big |_{\mu}
\ee 
The choice of parameter is natural if we recall that the expansion
inherent in the definition (\ref{laxdef}) must be carried out in 
descending powers of $\del$, which now corresponds to ascending powers
of $\mu$. 

The formulas (\ref{afflax}) simplify and unify after replacing
$\del$ by $1/\mu$: 
for each classical algebra and for ${\rm G}_2$, we find that 
\be
\widehat \La (u , 1/\mu) = \mu^{-h} \prod_{\lambda \in W} 
(1 + \mu \lambda \cdot u)
\ee
where $W$ is the set of weights of the defining representation.
Notice that, because the factors no longer involve operators, their ordering 
in the product is now irrelevant. Notice also that 
the overall power of $\mu$ corresponds to the fact that the order 
of $\widehat {\cal L}$ is always $h$. 
Substituting this expression into the previous formula above, 
and taking account of the overall power of $\mu$ that results, we 
find 
\begin{eqnarray}
h_{i_1 i_2 \ldots i_n} u^{i_1} u^{i_2} \ldots u^{i_n} & = &
\left ( \vphantom{A^2} \right . 
\prod_{\lambda \in W} 
(1 + \mu \lambda \cdot u) \, \left . \vphantom{A^2} 
\right )^{s/h} \Big |_{\mu^{s+1}} \nonumber \\
& = & \det ( \, 1 \, + \, \mu \, u^i t_i \, )^{s/h} \Big |_{\mu^{s+1}}
\end{eqnarray}
The last equality follows because the weights are,
by definition, the eigenvalues of the CSA generators (in 
the defining representation in this case).
This final formula for the $h$-tensors clearly coincides,
up to some irrelevant overall constants, with 
the definition (\ref{deteqn}) for the $k$-tensors when  
$j$ is restricted to the CSA.

\section{Symmetric space sigma-models}

The last topic we shall discuss is that of sigma-models on
compact symmetric spaces $G/H$. It was shown in \cite{EM} 
that commuting families of charges, with characteristic patterns of
spins, exist for each such model with $G$ and $H$ classical groups.
The approach we have developed in this paper is sufficient to 
extend the analysis to all symmetric space sigma-models,
and we now indicate briefly how this can be done  
(some routine details will be omitted in view of the strong 
similarities with the preceding discussions of Lie groups).

Most of the equations discussed above for 
PCM/WZW models carry over immediately to symmetric space 
sigma-models with appropriate re-interpretations of symbols.
For a symmetric space $G/H$, we have an orthogonal
decomposition of the Lie algebra $\g = \h + \p$, say.
The dynamical variables of the $G/H$ sigma-model are currents 
$j^a t_a$, where $\{ t_a \}$ is a basis for $\p$.
The conserved currents in this theory take the familiar form 
$ k_{a_1 \ldots a_m} j^{a_1} \ldots j^{a_m} $
but the symmetric $k$-tensor which appears must now be $H$-invariant on $\p$.
(In general, an $H$-invariant tensor $d$ on $\p$ satisfies (\ref{infinv})
for any $T \in \h$ and $U, V, \ldots , Z \in \p$.)
The condition for the corresponding conserved 
charges to commute is then still given by (\ref{kident}).

The analysis of \cite{EM} relied on the observation that when 
$G$ and $H$ are classical groups, every $H$-invariant tensor 
on $\p$ arises as the restriction of some $G$-invariant tensor on $\g$. 
This is not true for symmetric spaces involving 
exceptional groups, however \cite{Helg2} (see also \cite{Burstall},
where this point proved relevant).
To deal with these exceptional cases, we will follow 
the route developed in this paper for Lie groups: starting from 
Weyl-invariant tensors on a Cartan subalgebra and then extending 
them in an invariant fashion to construct the desired commuting charges.

A CSA for a compact symmetric space $G/H$ is 
a maximal set of commuting generators $\p_0 \subset \p$; this space 
is unique up to conjugation by elements of $H$ and the 
rank of $G/H$ is defined to be the dimension of $\p_0$.
Let $\{ t_i \}$ be a basis for $\p_0$, and $\{t_\alpha\}$
a basis for $\p_0^\perp$, its orthogonal complement in $\p$.
Any $X \in \p$ is conjugate by some $h \in H$ to a member of 
this CSA: $hXh^{-1} \in \p_0$ \cite{Helg}. 
The residual $H$-transformations which fix $\p_0$
constitute the Weyl group, which we shall denote ${\bf W}(G/H)$. 

As in the case of Lie groups, one can introduce the idea 
of a root system for a symmetric space,
and then summarize much of this information by means of a diagram 
which encodes the properties of a basis of simple roots.
It turns out that the diagram for any symmetric space $G/H$
coincides with the Dynkin diagram for some simple 
Lie group $K$, say \cite{Helg}.  
Moreover, rank($K$) = rank($G/H$) and ${\bf W}(K) = 
{\bf W}(G/H)$. We list the compact symmetric 
spaces $G/H$ and their diagrams, as given by $K$,
in appendix A.

Now, any $H$-invariant symmetric tensor on $\p$ is clearly 
determined by its restriction
to the CSA, $\p_0$. Furthermore, any tensor on $\p_0$ which is 
invariant under ${\bf W}(G/H)$ can be extended uniquely to an
$H$-invariant tensor on $\p$. 
A family of Weyl-invariant tensors 
$h_{i_1 \ldots i_m}$ on $\p_0$ is provided by the DS construction for 
the group $K$, and we can therefore extend these to $H$-invariant tensors
on $\p$ to define conserved currents in the $G/H$ sigma-model.
It remains to show that the conserved charges constructed 
in this manner really commute.
We know that the $h$-tensors satisfy (\ref{hident}), but we must 
promote this to the condition (\ref{kident}) on $\p$   
which means that we must generalize the restriction lemma of 
section 2 from Lie groups to symmetric spaces.

Following the same method as before, the proof 
of the restriction lemma will generalize to symmetric spaces
if each $H$-invariant tensor $d$ on $\p$ satisfies
\be\label{symlem}
d_{i_1 \ldots i_{n-1} \alpha} = 0 \qquad {\rm or} \qquad
d( X , \ldots , X , Y) = 0 \qquad {\rm for} \ X \in \p_0 \, , \
Y \in \p_0^\perp \ .
\ee
It is shown in \cite{Helg} (Chapter 7, Lemma 2.3) 
that one can choose pairs of generators 
$t_\alpha \in \p_0^\perp$ (the basis introduced above) and 
$s_\alpha \in \h$ such that, for each $X \in \p_0$,  
$[X ,t_\alpha ] = \lambda s_\alpha$ and 
$[X , s_\alpha ] = \lambda t_\alpha$ for some number 
$\lambda (X, \alpha)$.
Now fix $X$ and consider $Y = t_\alpha$. If $\lambda (X, \alpha) \neq 0$,
then (\ref{symlem}) follows from (\ref{infinv}) with 
$U = V = \ldots = Z = X$ and $T = s_\alpha$.
But if $\lambda (X, \alpha) = 0$, there exists 
some other $X' \in \p_0$ with $\lambda (X', \alpha) \neq 0$ (because
$\p_0$ is a CSA) and (\ref{symlem}) then follows from (\ref{infinv}) 
with $U = X'$, $V = \ldots = W = X$ and $T = s_\alpha$.
This completes the proof.

\section{Summary and Comments}

Using the results of Drinfeld and Sokolov, we have shown that 
there exist commuting charges in any PCM (or WZW model) based on a 
compact Lie group $G$, 
and that these charges have spins given by the exponents of the group 
modulo its Coxeter number.
We have also established analogous results for each sigma-model 
based on a compact symmetric space $G/H$, with the spins of the 
conserved charges given by the exponents of a related 
Lie group $K$ whose Dynkin diagram also encodes the root structure
of $G/H$ (see appendix A).
These results extend the work of \cite{EHMM1,EM}
for classical groups and symmetric spaces to include all the 
exceptional cases. 

Our construction involves a direct algebraic correspondence between 
the conserved, commuting charges in ATFTs, as given by 
the Drinfeld-Sokolov hierarchies, 
and those appearing in PCMs. 
Although this correspondence is formally very similar to the process 
of Hamiltonian reduction using gauged WZW models (reviewed briefly 
in section 4) we should be clear about how these procedures differ.
Unlike the well-known WZW-CTFT connection, our construction 
neither involves nor requires any dynamical relationship 
between PCMs (or WZW models) and ATFTs.
It is possible that some such relationship might be established,
by carrying out a non-conformal reduction of WZW models for instance.
Investigations of this kind have already been considered in the
literature, but their precise status remains rather unclear at 
present \cite{NCR}.

As part of our account, we have compared the detailed expressions
(\ref{laxdef}), given by Drinfeld and Sokolov for ATFT charges, with the
formula (\ref{deteqn}), originally introduced in \cite{EHMM1} for 
sigma-models, and found complete agreement in all cases where 
both are applicable.
This includes all conserved charges for models based on 
classical groups or on G${}_2$, with the exception of those charges   
associated with the Pfaffian invariant and its generalizations 
for the groups D${}_n$. In a sigma-model, the Pfaffian-type currents 
can still be extracted from (\ref{deteqn}) provided this formula 
is interpreted appropriately (see the account in \cite{EHMM1}) 
but there are apparently no explicit results for the corresponding 
Drinfeld-Sokolov or ATFT currents. 
For the other exceptional groups, beyond G${}_2$,
there are no known formulas of either type. It would be interesting
to rectify this.

One can regard sigma-models and Toda theories as two rather different
ways of introducing interactions amongst sets of free fields whilst
maintaining integrability. An important message which is already familiar
from hamiltonian reduction is that these two broad classes of models 
are much more closely related than might initially be supposed.
Our results reinforce this in a precise sense:
they reveal that the local commuting charges in both 
sigma-models (PCMs, WZW models, or symmetric space models)
and Toda theories (conformal or affine) 
are based on precisely the same sets of Weyl-invariant tensors.
\vskip 15pt

{\bf Acknowledgments}. I am grateful to Arthur Mountain, Niall Mackay 
and Ben Short for discussions; and to Fran Burstall and Alastair King 
for helpful remarks concerning references \cite{Helg2} 
and \cite{Burstall}. 
This work was carried out while on leave at Princeton, and 
I would like to thank the Physics Department as a whole 
and Erik Verlinde in particular for extending such a warm welcome 
during my stay. 
The research was funded by NSF grant PHY98-02484 and
by a PPARC Advanced Fellowship. 
\vfill \eject

\section*{Appendix A: Data for Lie groups and symmetric spaces}
\vskip 20pt
\centerline{\bf Table 1: Lie groups, exponents and Coxeter numbers}
\[
\begin{array}{cccc}
\hl
    \mbox{Lie group}~G~\mbox{or Lie algebra}~$\g$
& & \quad \mbox{exponents} \quad  & \mbox{Coxeter number}~h \quad \\ \hl
    {\rm A}_n = {\rm SU}(n{+}1) & & 1, 2, \dots , n & n{+}1 \\ \hl
    {\rm B}_n = {\rm SO}(2n{+}1) & & 1, 3, \dots , 2n{-}1 & 2n\\ \hl
    {\rm C}_n = {\rm Sp}(2n)  & & 1, 3, \dots , 2n{-}1 & 2n \\ \hl 
    {\rm D}_n = {\rm SO}(2n) & & 1, 3, \dots , 2n{-}3;n{-}1 
\quad & 2n{-}2\\ \hl
{\rm E}_6 & & 1, 4, 5, 7, 8, 11   & 12 \\ \hl 
{\rm E}_7 & & 1, 5, 7, 9, 11, 13, 17 & 18 \\ \hl 
{\rm E}_8 & & 1, 7, 11, 13, 17, 19, 23, 29 & 30 \\ \hl 
{\rm F}_4 & & 1, 5, 7, 11 & 12 \\ \hl 
{\rm G}_2 & & 1, 5        & 6 \\ \hl 
\end{array}
\]
\vfill \eject
\centerline{\bf Table 2(a): Symmetric spaces based on classical groups}
\[ 
\begin{array}{cccc}
\hl
    \mbox{Symmetric space}~G/H & 
       & \mbox{Simple root system of type $K$} \\ \hl
{\rm SU}(n{+}m) / {\rm S}({\rm U}(n){\times}{\rm U}(m)) \quad (n < m) 
& & {\rm B}_n \\ 
\\
{\rm SU}(2n)/{\rm S}({\rm U}(n){\times}{\rm U}(n)) 
& & {\rm C}_n \\ \hl
{\rm SO}(n{+}m)/{\rm SO}(n){\times}{\rm SO}(m) \quad (n < m) 
& & {\rm B}_n \\  
\\
{\rm SO}(2n)/{\rm SO}(n){\times}{\rm SO}(n) 
& & {\rm D}_n \\ \hl
{\rm Sp}(2n{+}2m)/{\rm Sp}(2n){\times}{\rm Sp}(2m) \quad (n < m) 
& & {\rm B}_n \\
\\ 
{\rm Sp}(4n)/{\rm Sp}(2n){\times}{\rm Sp}(2n)
& & {\rm C}_n \\ \hl
{\rm SU}(n)/{\rm SO}(n) 
& &{\rm A}_{n-1} \\ \hl
{\rm Sp}(2n)/{\rm U}(n) 
& & {\rm C}_n \\ \hl
{\rm SO}(4n)/{\rm U}(2n) 
& & {\rm C}_n \\ 
\\
{\rm SO}(4n+2)/{\rm U}(2n+1) 
& & {\rm B}_{n} \\ \hl
{\rm SU}(2n)/{\rm Sp}(2n) 
& & {\rm A}_{n-1} \\ \hl
  \end{array}
\]
\vfill \eject
\centerline{\bf Table 2(b): Symmetric spaces based on exceptional groups}
\[
\begin{array}{cccc}
\hl
    \mbox{Symmetric space}~G/H & 
       & \mbox{Simple root system of type}~K \\ \hl
{\rm E}_6 /{\rm Sp}(8)
& & {\rm E}_6 \\ \hl
{\rm E}_6 /{\rm SU}(6){\times}{\rm SU}(2)
& & {\rm F}_4 \\ \hl
{\rm E}_6 /{\rm SO}(10){\times}{\rm U}(1)
& & {\rm G}_2\\ \hl
{\rm E}_6 /{\rm F}_4 
& & {\rm A}_2  \\ \hl
{\rm E}_7/{\rm SU} (8) 
& & {\rm E}_7  \\ \hl
{\rm E}_7 /{\rm SO} (12){\times} {\rm SU}(2)
& & {\rm F}_4  \\ \hl
{\rm E}_7 /{\rm E}_6{\times}{\rm U}(1) 
& & {\rm C}_3  \\ \hl
{\rm E}_8 /{\rm SO} (16)  
& & {\rm E}_8  \\ \hl
{\rm E}_8 /{\rm E}_7{\times}{\rm SU}(2) 
& & {\rm F}_4  \\ \hl
{\rm F}_4 /{\rm Sp}(6){\times}{\rm SU}(2)
& & {\rm F}_4 \\ \hl
{\rm F}_4 /{\rm SO}(9)
& & {\rm A}_1 \\ \hl
{\rm G}_2 /{\rm SU}(2){\times}{\rm SU}(2)
& & {\rm G}_2 \\ \hl
  \end{array}
\]

\section*{Appendix B: Folding, ${\rm G}_2$, ${\rm SO(8)}$ and ${\rm
    F}_4$}

Consider a Lie algebra $\g$ and an automorphism $\sigma$ of order $n$.
There is a homomorphism $\pi : \g \rightarrow \g^\sigma$, the
$\sigma$-invariant subalgebra, given by $\pi (X) = 
X + \sigma (X) + \ldots + \sigma^{n-1} (X)$. 
Suppose $\sigma$ represents an outer automorphism and 
so corresponds to a non-trivial symmetry of the Dynkin diagram of
$\g$ with $n = 2$ or 3.
Identifying simple roots of $\g$ under this symmetry yields 
the Dynkin diagram for $\g^\sigma$, a process that is commonly
referred to as `folding' \cite{corri94}.
Applying this to simply-laced algebras of types A or D 
using their outer automorphisms of order 2 yields 
non-simply-laced algebras of types B or C.
Folding D${}_4$ = SO(8) using an automorphism of order 3 yields
${\rm G}_2$, the only exceptional group which can be constructed in
this fashion.

There is a one-to-one correspondence between (Ad)-invariant 
tensors on $\g^\sigma$ and (Ad)-invariant tensors on $\g$ which 
are additionally invariant under $\sigma$. In one direction 
this correspondence is given by restricting the tensor to the 
subalgebra (pulling back by the inclusion map) 
while in the other direction it is given by composing 
the tensor with the map $\pi$ in the obvious sense 
(pulling back using $\pi$).   
Furthermore, if we have a family of such $\sigma$-invariant
tensors $k^{(m)}$ on $\g$ which satisfy the key condition (\ref{kident}),
then it is not difficult to show (using arguments similar to those 
in the proof of the restriction lemma in section 2) that 
the corresponding tensors on $\g^\sigma$ satisfy (\ref{kident}) on 
this subalgebra, and vice versa. 

Now consider ${\rm G}_2$ obtained by folding SO(8) using 
$\sigma$ of order 3. Our calculations in section 3 
establish the existence of a family of tensors $k^{(m)}$ 
on ${\rm G}_2$ satisfying (\ref{kident}) with 
$m = 2, 6$ (mod 6). By the remarks above, these can also be 
regarded as $\sigma$-invariant tensors on SO(8).
(They do not coincide with the SO(8)-tensors defined 
by (\ref{deteqn}), however, which are not $\sigma$-invariant in general.)
But SO(8) is a subgroup of ${\rm F}_4$ of maximal rank;
moreover ${\rm F}_4$ is the minimal group in which 
the outer automorphisms of SO(8) become inner, and 
${\bf W}({\rm F}_4)$ is a semi-direct product of 
${\bf W}( {\rm SO(8)} )$ with the permutation group ${\rm S}_3$
of outer automorphisms \cite{Adams}. Since (Ad)-invariant tensors 
are determined by their values on a CSA, and the remnant of the
(Ad)-action on the CSA is the Weyl group, we see that the tensors 
$k^{(m)}$ on SO(8) can further be identified, via the common CSA, 
with (Ad)-invariant tensors 
on ${\rm F}_4$. The restriction lemma of section 2 ensures 
that these tensors still obey (\ref{kident}) and thus define commuting charges 
in the ${\rm F}_4$ model.
Notice also that the degrees of these tensors can be re-written as
$m = 2, 6, 8, 12$ (mod 12), exactly as expected.
This provides us with a convenient method for constructing the tensors
on ${\rm F}_4$ that we seek, although it does not alter the fact that these 
apparently cannot be derived by applying (\ref{deteqn}) directly to
${\rm F}_4$ itself.

The folding construction is widely used in Toda theory \cite{corri94}
and the outer automorphisms, or Dynkin diagram symmetries,
correspond directly to discrete symmetries of the Toda lagrangian.
The Lax operators of the simple form (\ref{conflax})
and (\ref{afflax}) for ${\rm G}_2$ can be derived directly 
from those for SO(8) by using this observation.
\vskip 25pt

\parskip=0pt

\end{document}